\documentclass[11pt]{article}
\usepackage{amsthm,amssymb,nath}
\def\CHANGE#1{#1}

\def\I{{\rm i}}

\def\eqref#1{{\rm(\ref{#1})}}
\def\Id{\mathop{\mathrm{Id}}\nolimits}
\def\R{\mathfrak R}
\def\hm{\hphantom{-}}
\newtheorem{proposition}{Proposition}
\newtheorem{corollary}{Corollary}
\newtheorem{definition}{Definition}
\newtheorem{procedure}{Procedure}
\newtheorem{remark}{Remark}
\newtheorem{example}{Example}

\newtheorem{conjecture}{Conjecture}

\begin{document}
\title{On the Spectral Parameter Problem}
\author{Michal Marvan}
\date{}

\maketitle

\begin{abstract}
We consider the problem whether a nonparametric zero-curvature representation can be embedded into a one-parameter family within the same Lie algebra.
After introducing a computable cohomological obstruction, \CHANGE{a method using the recursion operator to incorporate the parameter is discussed}. 
\end{abstract}

\section{Introduction}

Despite of four decades' development of soliton theory, a general algorithm to recognize integrability still remains elusive. 
By an integrable system in two independent variables $x,y$ we shall mean one that can be reproduced as the compatibility condition 
$$
\numbered\label{ZCR}
A_{y} - B_{x} + [A,B] = 0
$$
of the auxiliary linear system
$$
\numbered\label{lin problem}
\Psi_x = A \Psi, \quad \Psi_y = B \Psi.
$$
(\cite{Z-S,R-S}). Matrices $A,B$ are supposed to belong to a matrix Lie algebra $\mathfrak g$. The $\mathfrak g$-valued 1-form $A\,dx + B\,dy$ is called a {\it zero curvature representation}. To be indicative of integrability, matrices $A,B$ must depend on what is called the {\it spectral parameter}. 

Sometimes nonlinear systems come equipped with a nonparametric zero-curvature representation. In geometry of immersed surfaces, the Gauss--Mainardi--Codazzi equations are always of the form (\ref{ZCR}) while the linear problem (\ref{lin problem}) is supplied by the Gauss--Weingarten equations~\cite{Sym}. Integrable systems of this kind relevant to physics were first noticed by Lund and Regge~\cite{L-R}. For numerous classical and new examples see~\cite{Bob,R-S}. 
For a different source (the nonparametric Lax pairs~\cite{Lax}) see~\cite{L-T-L-F}.

Thus we \CHANGE{face} the ``spectral parameter problem'' or whether a given  nonparametric zero curvature representation can be embedded in a non-trivial one-parameter family. An arbitrary number of {\it removable} parameters can be always introduced through the gauge transformation 
$$
A' = S_x S^{-1} + S A S^{-1}, \qquad B' = S_y S^{-1} + S B S^{-1},
$$
where $S$ is an invertible matrix belonging to the Lie group $G$ associated with the Lie algebra $\mathfrak g$. The gauge transformation is a left action of the group of $G$-valued functions on the set of all solutions of system~\eqref{ZCR}. However, parameters introduced that way play no role in soliton theory. 

\CHANGE{In the geometric context, the majority of effective approaches link the spectral parameter with a deformation parameter (Bobenko~\cite{Bob}). 
A nonremovable parameter can be also often identified with the group parameter of a suitable point symmetry, see, e.g.,~\cite{Lund,Sas}. 
The early nineties~\cite{L-S-T,C-ls} witnessed a development of algorithmic procedures based on existing algorithms to compute symmetries. To detect integrable cases, Cie\'sli\'nski~\cite{C-ls} compares dimensions of the point symmetry algebras of the input system and of its covering~\cite{K-V} induced by the zero curvature representation. If these are inequal, a candidate for integrability is discovered.
Obviously, undetected remain the cases when the spectral parameter cannot be identified with a group parameter.
The first counterexample to be investigated thoroughly was that of inhomogeneous nonlinear Schr\"odinger system. This led Cie\'sli\'nski to consideration of extended symmetries~\cite{C-nls,C-nhnls} that operate within a class of equations. This method is suitable for classification problems, but results can depend on the initial class of equations considered (the extended symmetry must operate within this class).} 
\CHANGE{In this paper we present a general non-existence criterion for the spectral parameter problem within a given Lie algebra, a given jet order, and a given order of power expansion. Our main result can be viewed as a ``restricting'' procedure which says when a nonromovable parameter cannot be inserted. We also consider a new ``extending'' procedure that seeks the way to insert the parameter, thus trying to fill the space left by the restricting procedure.} The spectral parameter problem is solved (in certain jet order) when the results of restricting and extending procedures meet.

\section{Characteristic elements}
\label{sect:chare}

In this section we recall few necessary facts from the geometric theory of partial differential equations~\cite{B-V-V} and zero curvature representations~\cite{M1}.
Consider an arbitrary system of partial differential equations
$$
\numbered\label{sys}
E^l(x,y,u^i,u^i_x,u^i_t,\dots) = 0
$$ 
of an arbitrary order.
By $u^k_I$ we shall denote the derivative of $u^k$ specified by a symmetric multiindex $I$ in $x,y$. 
In the jet space with coordinates $t,x,u^k_I$ we consider {\it total derivatives}
$$
D_x = \frac\partial{\partial x}
 + \sum_{i,I} u_{xI} \frac{\partial}{\partial u^i_I}, \qquad
D_t = \frac\partial{\partial t}
 + \sum_{i,I} u_{tI} \frac{\partial}{\partial u^i_I}.
$$ 
Equations~\eqref{sys} and their differential consequences $D_I E^l = 0$ determine a submanifold (called {\it diffiety}), which we shall denote by $\mathcal E$. Here $D_I$ denotes the composition of the operators of total derivative $D_x$, $D_y$ (in any order, since they commute). Finally, operators $D_x$, $D_y$ admit restriction to the diffiety.

A zero curvature representation is defined to be a $1$-form $\alpha = A\,dx + B\,dy$ such that
$$
\numbered\label{ZCR'}
D_{y} A - D_{x} B + [A,B] = 0
$$
on $\mathcal E$.
Two zero curvature representations $A\,dx + B\,dy$ and $A'\,dx + B'\,dy$ are said to be {\it gauge equivalent} if one can be obtained from the other by the gauge transformation
$$
A' = D_x S\,S^{-1} + S A S^{-1}, \qquad B' = D_y S\,S^{-1} + S B S^{-1}.
$$ 
The gauge equivalence explains why system~\eqref{ZCR'} is so much underdetermined (has twice as many unknowns as equations), which is the main obstacle to successful search for its solutions. 
To compute zero curvature representations, we can choose between the classical Wahlquist--Estabrook method of prolongation algebras~\cite{W-E,D-F} (see \cite{Gra} for implementation), its extensions such as~\cite{F-M}, the method of coverings by Krasil'shchik and Vinogradov~\cite{K-V}, the remarkable development by Igonin~\cite{Igo}, the Sakovich method of cyclic bases~\cite{Sak0,Sak}, and also the method of characteristic elements~\cite{M1,M2}. Being a modification of the well-known characteristic function method to compute conservation laws, the latter approach tries to fix the gauge and copes with classification problems reasonably well. 

Given a zero curvature representation $A\,dx + B\,dy$ of system~\eqref{sys}, the characteristic elements can be computed using the Sakovich formula~\cite{Sak0}. Namely, we necessarily have 
$D_y A - D_x B + [A,B] = \sum_{J,l} C^J_l D_J E^l$ with suitable matrices $C^J_l$, and then the $l$th characteristic element $C_{(l)}$ is given by
$$
\numbered\label{Ch}
C_{(l)} = \left.\sum_J (-\hat D)_J C^J_l\right|_{\mathcal E}.
$$ 
Here $\hat D_x = D_x - [A,\adot]$, $\hat D_y = D_y - [B,\adot]$, and $\hat D_I$ is their obvious composition (in any order since $\hat D_x, \hat D_y$ necessarily commute once $A,B$ form a zero curvature representation).
Finally, $|_{\mathcal E}$ denotes the restriction to $\mathcal E$.

\begin{example}\rm
\label{mKdV}
The mKdV equation $u_t + u_{xxx} - 6 u^2 u_x = 0$
has an $\mathfrak{sl}(2)$-valued zero curvature representation $A\,dx + B\,dt$ with
$$
A = (\begin{array}{cc}
u & \hphantom{-}\lambda \\ 1 & -u
\end{array}\biggr),
\quad
B = \biggl(\begin{array}{cc}
-u_{xx} + 2 u^3 - 4 \lambda u & 
2 \lambda u_x + 2 \lambda u^2 - 4 \lambda^2 \\
-2 u_x + 2 u^2 - 4 \lambda & 
u_{xx} - 2 u^3 + 4 \lambda u
\end{array}).
$$
Actually, we have $D_t(A) - D_x(B) + [A,B]
 = (u_t + u_{xxx} - 6 u^2 u_x) \cdot C_{(1)}$,
where 
$$
C_{(1)} = (\begin{array}{cr} 1 & 0 \\ 0 & -1 \end{array}).
$$ 
By formula~\eqref{Ch}, $C_{(1)}$ is the characteristic element. The only nonzero summand in~\eqref{Ch} corresponds to \CHANGE{the} empty multiindex $J$. This is typical for zero curvature representations computable by the Wahlquist--Estabrook procedure.
\end{example}

Under gauge transformation, characteristic elements transform by conjugation, which allows us to bring one of them into the Jordan normal form~$J$.
This usually leaves some gauge freedom, which can be exploited to reduce one of the matrices $A_i$ by gauge transformation with respect to the \CHANGE{relatively small} stabilizer subgroup $\mathcal S_J$ of $J$. This very much resembles the well-known ``wild'' problem of normal forms of pairs of matrices with respect to simultaneous conjugation. In reality \CHANGE{our} problem is simpler, since zero curvature representations of use in soliton theory never take values in a resolvable subalgebra, which allows us to cut branches that fall into such a subalgebra.  
In the case of Lie algebra $\mathfrak{sl}(2)$ and $\mathfrak{sl}(3)$ all the normal forms were found in~\cite{M2} and~\cite{Seb3}, respectively.
For convenience we tabulate them in the two columns headed $k = 0$ of Table~\ref{tabsl2} and~\ref{tabsl3}. 
The remaining columns, headed $k \gt 0$, will be explained in Section~\ref{H1}.

\begin{table}
\caption{Normal forms in the case of algebra \protect$\mathfrak{sl}(2)\protect$}
\label{tabsl2}
$$
\begin{array}{cccc}
\multicolumn{2}{c}{C} & \multicolumn{2}{c}{A}  
\\\hline
k = 0 & k \gt 0 & k = 0 & k \gt 0
\\\hline\hline
(\begin{array}{cr} c & 0 \\ 0 & -c \end{array}) &
(\begin{array}{cr} c & 0 \\ 0 & -c \end{array}) &
(\begin{array}{cc} a_{11} & \hm a_{12} \\ 1 & -a_{11} \end{array}) &
(\begin{array}{cc} a_{11} & \hm a_{12} \\ 0 & -a_{11} \end{array}) 
\\\hline
(\begin{array}{cc} 0 & 1 \\ 0 & 0 \end{array}) &
(\begin{array}{cc} 0 & 0 \\ c & 0 \end{array}) &
(\begin{array}{cc} 0 & a_{12} \\ a_{21} & 0 \end{array}) &
(\begin{array}{cc} 0 & a_{12} \\ a_{21} & 0 \end{array}) 
\end{array}
$$
\end{table}

\begin{table}
\caption{Normal forms in the case of algebra \protect$\mathfrak{sl}(3)\protect$}
\label{tabsl3}
$$
\begin{array}{cccc}
\multicolumn{2}{c}{C} & \multicolumn{2}{c}{A}  
\\\hline
k = 0 & k \gt 0 & k = 0 & k \gt 0
\\\hline\hline
(\begin{array}{ccc} c_1 & 0 & 0 \\ 0 & c_2 & 0 \\
 0 & 0 & c_3 \end{array}) &
(\begin{array}{ccc} c_1 & 0 & 0 \\ 0 & c_2 & 0 \\
 0 & 0 & c_3 \end{array}) &
(\begin{array}{ccc} a_{11} & a_{12} & a_{13} \\ 1 & a_{22} & a_{23} \\
 a_{31} & 1 & a_{33}\end{array}) &
(\begin{array}{ccc} a_{11} & a_{12} & a_{13} \\ 0 & a_{22} & a_{23} \\
 a_{31} & 0 & a_{33}\end{array}) 
\\\hline
(\begin{array}{ccc} c & 0 & \hm 0 \\ 0 & c & \hm 0 \\
 0 & 0 & -2 c \end{array}) &
(\begin{array}{ccc} c & 0 & \hm 0 \\ 0 & c & \hm 0 \\
 0 & 0 & -2 c \end{array}) &
(\begin{array}{ccc} 0 & 1 & 0 \\ a_{21} & a_{22} & 1 \\
 a_{31} & a_{32} & a_{33}\end{array}) &
(\begin{array}{ccc} 0 & 0 & 0 \\ a_{21} & a_{22} & 0 \\
 a_{31} & a_{32} & a_{33}\end{array}) 
\\\hline
(\begin{array}{ccc} c & 0 & \hm 0 \\ 1 & c & \hm 0 \\
 0 & 0 & -2 c \end{array}) &
(\begin{array}{ccc} c & c_1 & \hm 0 \\ 0 & c & \hm 0 \\
 0 & 0 & -2 c \end{array}) &
\begin{array}{c}
 (\begin{array}{ccc} a_{11} & a_{12} & 0 \\ a_{21} & a_{22} & 1 \\
  a_{31} & a_{32} & a_{33} \end{array}) \\
 (\begin{array}{ccc} 0 & a_{12} & 0 \\ a_{21} & a_{22} & 1 \\
  a_{31} & 0 & a_{33} \end{array}) 
\end{array} & 
\begin{array}{c}
 (\begin{array}{ccc} a_{11} & a_{12} & 0 \\ a_{21} & a_{22} & 0 \\
  a_{31} & a_{32} & a_{33} \end{array}) \\
 (\begin{array}{ccc} 0 & a_{12} & a_{13} \\ a_{21} & a_{22} & 0 \\
  a_{31} & a_{32} & a_{33} \end{array}) 
\end{array}
\\\hline
(\begin{array}{ccc} 0 & 0 & 0 \\ 1 & 0 & 0 \\ 0 & 0 & 0 \end{array}) &
(\begin{array}{ccc} c & c_{12} & \hm c_{13} \\ 0 & c & \hm 0 \\
 0 & c_{32} & -2 c \end{array}) &
\begin{array}{c}
 (\begin{array}{ccc} 0 & a_{12} & 0 \\ a_{21} & a_{22} & 1 \\
  a_{31} & 0 & a_{33} \end{array}) \\
 (\begin{array}{ccc} 0 & 0 & 1 \\ a_{21} & 0 & 0 \\ a_{31} & 0 & 0\end{array})
\end{array} &
\begin{array}{c}
 (\begin{array}{ccc} 0 & a_{12} & 0 \\ a_{21} & a_{22} & 0 \\
  a_{31} & 0 & a_{33} \end{array}) \\
 (\begin{array}{ccc} 0 & a_{12} & 0 \\ a_{21} & 0 & 0 \\
 a_{31} & a_{32} & 0 \end{array})
\end{array} 
\\\hline
(\begin{array}{ccc} 0 & 0 & 0 \\ 1 & 0 & 0 \\ 0 & 1 & 0 \end{array}) &
(\begin{array}{ccc} 0 & c_1 & c_2 \\ 0 & 0 & c_1 \\ 0 & 0 & 0 \end{array}) &
\begin{array}{c}
 (\begin{array}{ccc} a_{11} & a_{12} & 1 \\ a_{21} & a_{22} & 0 \\
  a_{31} & a_{32} & a_{33} \end{array}) \\
 (\begin{array}{ccc} 0 & a_{12} & 0 \\ a_{21} & a_{22} & a_{23} \\
 a_{31} & 0 & a_{33} \end{array})
\end{array} &
\begin{array}{c}
 (\begin{array}{ccc} 0 & 0 & a_{13} \\ a_{21} & a_{22} & a_{23} \\
  a_{31} & a_{32} & a_{33} \end{array}) \\
 (\begin{array}{ccc} 0 & a_{12} & a_{13} \\ a_{21} & a_{22} & a_{23} \\
  a_{31} & 0 & a_{33} \end{array}) 
\end{array}
\end{array}
$$
All matrices are supposed to be traceless (e.g., $c_3 = -c_1 - c_2$).
\end{table}

The general case of Lie algebra $\mathfrak{sl}(n)$ still remains largely unexplored except when the Jordan normal form $J$ consists of a single Jordan block, which has been completely investigated by Sebesty\'en~\cite{Sebn}.  

All possible zero curvature representations \eqref{ZCR'} along with their characteristic elements~\eqref{Ch} now can be found from the determining system
$$
\numbered\label{ds}
\left. D_{y} A - D_{x} B + [A,B] \right|_{\mathcal E} = 0, \\
\left.
 \sum_{I,l} (-\hat D)_I (\frac{\partial E^l}{\partial u^k_I} C_{(l)})
\right|_{\mathcal E} = 0, 
$$
derived in~\cite{M1}.
With $A,B,C_{(l)}$ restricted to their respective normal forms, the determining system~\eqref{ds} enjoys the following properties:

\paritem{--} is a system of differential equations in total derivatives;
\paritem{--} has the same number of unknowns as equations, akin to systems determining symmetries and conservation laws;
\paritem{--} is quasilinear in $A,B$, and linear in $C_{(l)}$.

Solution of system \eqref{ds} is essentially algorithmic given an upper bound on the jet order of the unknowns. Because of nonlinearity in $A$ and $B$, solution can be troublesome even with the help of computer algebra.

\begin{remark} \rm
\label{W-E}
Wahlquist--Estabrook type zero curvature representations have the order equal to the order of the equation minus one. The order of their normal form can be twice the order of the equation minus one at worst. 
\end{remark}

Let us remark that the procedure does not need to fix the gauge completely. The general solution of system~\eqref{ds} may still depend on a removable ``false'' parameter.
The parameter originates from linearity of system~\eqref{ds} in the characteristic elements as demonstarted in the following example.

\begin{example} \rm
\label{false0}
Consider the case of the Lie algebra $\mathfrak{sl}(2)$ and the Jordan normal form
$$
J = (\begin{array}{cc}
0 & 0 \\ 1 & 0
\end{array})$$
of the characteristic element. The 1-parameter gauge group
$$
\numbered\label{false_sl2}
S = (\begin{array}{cc}
s & 0 \\ 0 & s^{-1}
\end{array}), \qquad s \ne 0,
$$
preserves the normal form of $A$ and multiplies $J$ by $s^{-2}$. 
By linearity, if $A,B,C_{(l)}$ is a solution of system~\eqref{ds}, then 
$A,B,s^2 C_{(l)}$ is a solution as well.
If $C_{(1)} = J$ and $A$ is normalized according to Table~\ref{tabsl2}, then $A,B,s^2 C_{(l)}$ is an unnormalized solution, but $S A S^{-1}$, $S B S^{-1}$, $S s^2 C_{(l)} S^{-1}$ is a normalized solution.
Thus~\eqref{false_sl2} introduces one false parameter into the solution of system~\eqref{ds} under normalization.
\end{example}

So far, the only complete classification (with unbounded jet order) obtained is~\cite{ssoee} (second order evolution equations possessing a $\mathfrak{sl}(2)$-valued zero curvature representations; none admitting an essential parameter, hence all nonintegrable as naturally expected).

\section{Restriction procedure. Formal removability}
\label{H1}

In the previous work~\cite{hgc}, a nontrivial horizontal gauge cohomology group $H^1$ was associated with zero curvature representations depending on a nonremovable parameter. Here we elaborate the idea further in terms of power series in the spectral parameter. 

Given a nonparametric zero curvature representation $A_0, B_0$, consider all possible expansions
$$
\numbered\label{TaylorAB}
A(\lambda) = \sum_{i = 0}^\infty A_i \lambda^i, \quad
B(\lambda) = \sum_{i = 0}^\infty B_i \lambda^i.
$$
around zero.
Obviously, $A_0 = A(0)$, $B_0 = B(0)$. 

Upon inserting expansions \eqref{TaylorAB} into the zero curvature condition~\eqref{ZCR'} we obtain
$$
\numbered\label{ABk}
D_{y} A_k - D_{x} B_k + \sum_{i + j = k} [A_i,B_j] = 0
$$
for all $k \ge 0$.
For $k = 0$ this equation says just that $A_0,B_0$ is a zero curvature representation.
For $k = 1$, equation \eqref{ABk} reads
$$
\numbered\label{AB1}
D_{y} A_1 - D_{x} B_1 + [A_1,B_0] + [A_0,B_1] = 0
$$
\CHANGE{(according to~\cite{F-G}, this equation also plays an important role in the geometry of surfaces immersed in Lie algebras).}
Denoting $\hat D_x = D_x - [A_0,\adot]$, $\hat D_y = D_y - [B_0,\adot]$,  equation \eqref{AB1} can be rewritten as 
$$
\numbered\label{2nd}
\hat D_y A_1 = \hat D_x B_1
$$ 
and interpreted cohomologically~\cite{hgc}.

Recall that the horizontal gauge complex~\cite{hgc} consists of modules $\mathfrak g \otimes \bar\Lambda^k$ of horizontal forms with coefficients in the Lie algebra $\mathfrak g$. Here horizontal essentially means that the forms contain differentials $dx, dy$ only and $\bar\Lambda^0$ is simply the additive group of functions. The operator $\hat d$ acts on an arbitrary $\mathfrak g$-valued function $F \in \mathfrak g \otimes \bar\Lambda^0$ as 
$$
\hat d F = \hat D_x F\,dx + \hat D_y F\,dy \in \mathfrak g \otimes \bar\Lambda^1,
$$ 
while for an arbitrary $\mathfrak g$-valued horizontal $1$-form $U\,dx + V\,dy \in \mathfrak g \otimes \bar\Lambda^1$ we have
$$
\hat d(U\,dx + V\,dy)
 = (\hat D_x V - \hat D_y U)\,dx \wedge dy \in \mathfrak g \otimes \bar\Lambda^2.
$$ 
It is easy to see that $\hat d \circ \hat d$ = 0 once $A_0,B_0$ is a zero curvature representation.
The group
$$
H^1 = \frac{`Ker \hat d : g \otimes \bar\Lambda^1 \to g \otimes \bar\Lambda^2}
  {`Im \hat d : g \otimes \bar\Lambda^0 \to g \otimes \bar\Lambda^1}
$$
is called the first horizontal gauge cohomology group of the zero curvature representation $A_0,B_0$. Elements of $`Ker \hat d$ are called cocycles while those of $`Im \hat d$ are coboundaries.

We show that $H^1 = 0$ implies removability of the parameter in a strict sense.
To circumvent the issue of convergence, we adopt the following definition.

\begin{definition} \rm
\label{def:frem}
The parameter $\lambda$ is said to be {\it formally removable} if every summand in the expansions~\eqref{TaylorAB} except $A_0,B_0$ can be locally annihilated by a gauge transformation with respect to some matrix function in $\lambda$. 
\end{definition}

The following proposition is easy to prove.

\begin{proposition}
\label{prop:frem}
Let $A(\lambda),B(\lambda)$ be a zero curvature representation depending analytically on a parameter $\lambda$. Let the first horizontal gauge cohomology group $H^1$ with respect to $A_0,B_0$ be zero. Then $\lambda$ is formally removable. 
\end{proposition}

\begin{proof}
Equation \eqref{2nd} says that $A_1\,dx + B_1\,dy$ is a cocycle, hence also a coboundary if $H^1 = 0$. This means that $A_1\,dx + B_1\,dy = \hat d S_1$ for a suitable $\mathfrak g$-valued matrix function $S_1$, i.e.,
$$
A_1 = D_x S_1 - [A_0,S_1], \\
B_1 = D_y S_1 - [B_0,S_1].
$$
Let $S(\lambda) = E + S_1 \lambda$. The gauge action with respect to $S(\lambda)$ will keep the terms $A_0,B_0$, but remove the terms $A_1,B_1$. This is readily checked using the expansion $(E + S_1 \lambda)^{-1} = E - S_1 \lambda  + S_1^2 \lambda^2 - S_1^3 \lambda^3 + \dots$, which has a positive radius of convergence in every open set where $E + S_1$ has bounded eigenvalues, see Gantmacher~\cite[Ch.~V, \S~4, Thm.~2]{Gant}. This means that $A_1,B_1$ can be locally annihilated.
Otherwise said, if $H^1 = 0$, then without loss of generality we can assume that $A_1 = B_1 = 0$.

Now we can proceed by induction. If $A_i = B_i = 0$ for all $i < k$, then
equation \eqref{ABk} with $k \ge 2$ reads
$$
D_{y} A_k - D_{x} B_k + [A_k,B_0] + [A_0,B_k] = 0,
$$
meaning that $A_k\,dx + B_k\,dy$ is a cocycle, hence a coboundary, i.e., 
equal to $\hat d S_1$ for suitable $S_k$.
Reasoning as above, one easily sees that $A_k,B_k$ can be locally gauged out through the gauge matrix $E + S_k \lambda^k$. 
This concludes the proof.
\end{proof}

\begin{corollary}
\label{prop:h1}
A zero curvature representation $A_0,B_0$ such that the first horizontal gauge cohomology group $H^1$ is zero cannot be a member of an analytic one-parameter family that would depend on a formally nonremovable parameter. 
\end{corollary}

A method to compute the cohomology group $H^1$ of a given zero curvature representation $A_0,B_0$ follows from~\cite[Prop.~7]{hgc}, which we reproduce below in a slightly simplified form:

\begin{proposition}[\cite{hgc}]
Given a zero curvature representation $A_0\,dx + B_0\,dy$, the $1$-form $A_1\,dx + B_1\,dy$ is a cocycle if and only if matrices
$$
\numbered\label{bar AB}
A^{[1]} = (\begin{array}{cc} A_0 & 0 \\ 
A_1 & A_0 \end{array}),
\qquad
B^{[1]} = (\begin{array}{cc} B_0 & 0 \\ 
B_1 & B_0 \end{array}),
$$
constitute a zero curvature representation:
$$
\numbered\label{bar zcr}
D_{y} A^{[1]} - D_{x} B^{[1]} + [A^{[1]},B^{[1]}] = 0.
$$
Moreover, two cocycles are cohomological (differ by a coboundary) if and only if the corresponding zero curvature representations~\eqref{bar AB} are gauge equivalent with respect to a gauge matrix of the block triangular form
$$
\numbered\label{bar S}
S^{[1]} = (\begin{array}{cc} E & 0 \\ S & E \end{array})
$$
with the unit matrix $E$ at the diagonal positions.
\end{proposition}

It follows that cohomology classes of cocycles $A_1\,dx + B_1\,dy$ can be determined by solving equation~\eqref{bar zcr} modulo gauge equivalence with respect to~\eqref{bar S}. 
If $A_1 = 0$, $B_1 = 0$ for all \CHANGE{normalized} solutions to~\eqref{bar zcr}, then $H^1 = 0$.
However, there can also be a kind of ``false'' solutions to equation~\eqref{bar zcr}.

\begin{example} \rm
\label{false1}
Continuing Example~\ref{false0}, consider the gauge matrix
$$
S = (\begin{array}{cc} 1 + c_1 \lambda + c_2 \lambda^2 + \dots & 0 \\
 0 & (1 + c_1 \lambda + c_2 \lambda^2 + \dots)^{-1} \end{array})
\\\quad
 = (\begin{array}{cc} 1 + c_1 \lambda + c_2 \lambda^2 + \dots & 0 \\
 0 & 1 - c_1 \lambda + (c_1^2 - c_2) \lambda^2 + \dots \end{array})
$$
with constant coefficients $c_i$.
In Example~\ref{false0} we already demonstrated that $S$ introduces a false parameter not detected by the characteristic element method.
What is then the action of $S$ on the matrix 
$A_0 = \displayed{(\begin{array}{rr} a_{11} & a_{12} \\ a_{21} & -a_{11} \end{array})}$?
By expanding $S A_0 S^{-1}$ in the powers of $\lambda$ we get 
$$
S A_0 S^{-1} = A_0
 + (\begin{array}{cc} 0 & -2 c_1 a_{12} \\
      2 c_1 a_{21} & 0 
   \end{array}) \lambda
\\\qquad + (\begin{array}{cc} 0 & (3 c_1^2 - 2 c_2) a_{12} \\
      (c_1^2 + 2 c_2) a_{21} & 0 
   \end{array}) \lambda^2
 + \dots.
$$
In particular, $A_1 = \displayed{(\begin{array}{cc} 0 & -2 c_1 a_{12} \\ 2 c_1 a_{21} & 0 \end{array})}$. If this $A_1$ exhausts solutions to~\eqref{bar zcr}, then $H^1 = 0$ again.  
\end{example}

If $H^1 = 0$, then Proposition~\ref{prop:frem} applies. 
Otherwise, $A_1,B_1$ are candidates for the first coefficients of the Taylor expansions~\eqref{TaylorAB}. How to check whether they are good?

It is easily verified that the system of equations~\eqref{ABk} indexed by $k = 0,\dots,m$ is satisfied if and only if the block triangular matrices
$$
\numbered\label{barbar AB}
A^{[m]} = (\begin{array}{cccc} A_0 & 0 & \dots & 0 \\ 
A_1 & A_0 & \ddots & \vdots \\
\vdots & \ddots & \ddots & 0 \\
A_m & \dots & A_1 & A_0 \end{array}),
\quad
B^{[m]} = (\begin{array}{cccc} B_0 & 0 & \dots & 0 \\ 
B_1 & B_0 & \ddots & \vdots \\
\vdots & \ddots & \ddots & 0 \\
B_m & \dots & B_1 & B_0 \end{array}),
$$
constitute a zero curvature representation, i.e.,
$$
\numbered\label{barbar zcr}
D_{y} A^{[m]} - D_{x} B^{[m]} + [A^{[m]},B^{[m]}] = 0. 
$$
Hence, if $A_k,B_k$ are already known for all $k < m$, then~\eqref{barbar zcr} \CHANGE{determines} whether expansions~\eqref{TaylorAB} can be extended one step further.

\begin{proposition}
\label{prop:ext}
If, for some $m$, no solution $A^{[m]},B^{[m]}$ of equation~\eqref{barbar zcr} exists, then there is no possibility to extend the expansions~\eqref{TaylorAB} beyond the first $m$ terms.
\end{proposition}

Propositions~\ref{prop:h1} and~\ref{prop:ext} reduce the spectral parameter problem to that of computation of zero curvature representations modulo a gauge group. The substantial benefit is that the problem becomes linear, being reducible to solution of a system~\eqref{ds} written on matrices $A^{[1]}$, $B^{[1]}$ given by~\eqref{bar AB}, resp. on matrices $A^{[m]}$, $B^{[m]}$ given by~\eqref{barbar AB}. 
The system~\eqref{ds} is then indeed linear in all unknowns including $A_k,B_k$, $k \ge 1$ (being still nonlinear in $A_0,B_0$, which are supposed to be known, however).

To solve system~\eqref{bar zcr} or~\eqref{barbar zcr} by the characteristic element method, we need to construct the respective normal forms. 
The characteristic elements of the zero curvature representation $A^{[k]}, B^{[k]}$ are block triangular matrices of the same form
$$
\numbered\label{barbar C}
C_{(l)}^{[k]} = (\begin{array}{cccc} C_{(l)0} & 0 & \dots & 0 \\ 
C_{(l)1} & C_{(l)0} & \ddots & \vdots \\
\vdots & \ddots & \ddots & 0 \\
C_{(l)k} & \dots & C_{(l)1} & C_{(l)0} \end{array}).
$$ 
Every diagonal block $C_{(l)0}$ coincides with the corresponding known characteristic element of the input zero curvature representation $A_0,B_0$.
The gauge transformations considered should be with respect to matrices of the form
$$
S^{[k]} = (\begin{array}{cccc} 
E & 0 & \dots & 0 \\ 
S_1 & E & \ddots & \vdots \\
\vdots & \ddots & \ddots & 0 \\
S_k & \dots & S_1 & E \end{array}).
$$ 
In the case of the Lie algebras $\mathfrak{sl}(2)$ and $\mathfrak{sl}(3)$ the normal forms for $C_k$ and $A_k$, $k > 0$, are tabelated in Table~\ref{tabsl2} and~\ref{tabsl3} above. 

Being a determined system of equations in total derivatives, system~\eqref{ds} is routinely solvable if one sets an upper bound on the jet order of the unknowns (see the end of Section~\ref{sect:chare}). It seems natural to assume that insertion of the paramater does not require increasing the jet order of the zero curvature representation. However, this is not necessarily the same thing as setting the order of $A_k,B_k$ equal to that of $A_0,B_0$, since normalization can increase the jet order (see Remark~\ref{W-E} and the example in the next section).

\section{An example. The inhomogeneous nonlinear Schr\"o\-dinger equation}
\label{sect:nhnls1}

In this section we consider Cie\'sli\'nski's counterexample~\cite{Bal,C-nls,C-nhnls,C-G-S}, the inhomogeneous nonlinear Schr\"odinger equation 
\begin{eqnarray}
q_t &=&  (f q)_{xx} + 2 q r, \nonumber
\\
p_t &=&  -(f p)_{xx} - 2 p r, \label{nhnls}
\\
r_x &=&  f (p q)_x + 2 f_x p q, \nonumber
\end{eqnarray}
where $f(t,x)$ is an arbitrary function. Compared to Cie\'sli\'nski's formulation~\cite{C-nls,C-nhnls,C-G-S}, we have put $p = \bar q$ and made the formal change of coordinates $t \leftrightarrow \I t$, preserving integrability. The system is known to be integrable 
if $f = f_1(t) x + f_0(t)$ is linear in $x$; otherwise it is believed to be nonintegrable (see op. cit. and references therein).

The initial nonparametric first order $\mathfrak{sl}(2)$-valued zero curvature representation is 
$D_t A_0 - D_x B_0 + [A_0,B_0] = 0$ with
$$
\numbered\label{nhnls AB}
A_0 = 
\left(\begin{array}{rc}
0 & q \\ -p & 0
\end{array}\right), \quad
B_0 = \left(\begin{array}{cc}
r & (f q)_x \\ (f p)_x & -r
\end{array}\right),
$$
valid for all functions $f(t,x)$. 
  
Let us demonstrate how the procedure outlined in the previous section detects the expected integrable cases in two steps. To start with, we find the characteristic element of $A_0\,dx + B_0\,dy$ to be the triple of $\mathfrak{sl}(2)$-matrices
$$
C_{(1)0} = (\begin{array}{rc} 0 & 1 \\ 0 & 0 \end{array}), \quad
C_{(2)0} = (\begin{array}{rc} 0 & 0 \\ -1 & 0 \end{array}), \quad
C_{(3)0} = (\begin{array}{rc} -1 & 0 \\ 0 & 1 \end{array}).
$$
Indeed, the expression $D_t A_0 - D_x B_0 + [A_0,B_0]$ is equal to the sum of $C_{(j)0}$ multiplied by $j$th equation~\eqref{nhnls} each, similarly as in Example~\ref{mKdV} above.
  
To proceed further, we construct the $4 \times 4$ matrices~\eqref{bar AB} and~\eqref{barbar C}, $k = 1$, from the already known blocks
$A_0,B_0,C_{(1)0},C_{(2)0},C_{(3)0}$ and the blocks $A_1,B_1,C_{(1)1},C_{(2)1},C_{(3)1}$ yet to be found.
\CHANGE{These matrices are subject to system~\eqref{ds}.
At this point there are essentially two natural} ways to proceed further depending on which pair of the unknown matrices we choose to normalize.

{\it Choice 1. } According to the second row of Table~\ref{tabsl2} above, matrices $C_{(1)0}$ and $A_0$ are already in their respective normal forms. However, $C_{(1)0}$ is a single Jordan block, hence a ``false'' parameter should be expected (see Example~\ref{false0}).

According to Table~\ref{tabsl2} above, normal forms for $C_{(1)1}$ and $A_1$ are
$$
C_{(1)1} = (\begin{array}{cc} 0 & 0 \\ c & 0 \end{array}), \quad
A_1 = (\begin{array}{cc} 0 & a_{12} \\ a_{21} & 0 \end{array}).
$$
Then there is no room for further normalization and matrices  $B_1,C_{(2)1},C_{(3)1}$ have to be left arbitrary. Therefore, the $4 \times 4$ matrices 
$$
A^{[1]} = (\begin{array}{cc@{\;\quad}cc}
0 & q & 0 & 0 \\
\llap{$-$}p & 0 & 0 & 0 \\
0 & a_{12} & 0 & q \\
a_{21} & 0 & \llap{$-$}p & 0
\end{array}), \\
B^{[1]} = (\begin{array}{cccc}
 r & (f q)_x & 0 & 0 \\
(f p)_x & \llap{$-$}r & 0 & 0 \\
b_{11} & b_{12} & r & (f q)_x \\
b_{21} & \llap{$-$}b_{11} & (f p)_x & \llap{$-$}r  
\end{array}), \\
C_{(1)}^{[1]} = (\begin{array}{cccc}
0 & 1 & 0 & 0 \\
0 & 0 & 0 & 0 \\
0 & 0 & 0 & 1 \\
c & 0 & 0 & 0 
\end{array}), \\
C_{(2)}^{[1]} = (\begin{array}{c@{\;\quad}c@{\;\quad}cc}
0 & 0 & 0 & 0 \\
\llap{$-$}1 & 0 & 0 & 0 \\
c_{(2)11} & c_{(2)12} & 0 & 0 \\
c_{(2)21} & \llap{$-$}c_{(2)11} & \llap{$-$}1 & 0 
\end{array}), \\
C_{(3)}^{[1]} = (\begin{array}{c@{\;\quad}c@{\;\quad}cc}
\llap{$-$}1 & 0 & 0 & 0 \\
 0 & 1 & 0 & 0 \\
c_{(3)11} & c_{(3)12} & \llap{$-$}1 & 0 \\
c_{(3)21} & \llap{$-$}c_{(3)11} & 0 & 1 
\end{array})
$$ 
involve $12$ unknown functions, namely $a_{11},a_{12},b_{11},b_{12},b_{21},c$, $c_{(2)11},c_{(2)12},c_{(2)21}$, $c_{(3)11},c_{(3)12},c_{(3)21}$.
The determining system~\eqref{ds} consists of the same number 12 of independent linear equations in total derivatives. 
Assuming $A_1,B_1$ of the second order at least (cf. Remark~\ref{W-E}), we obtain a nontrivial solution
$$
\numbered\label{nhnls H^1}
A_1 = (\begin{array}{cr} 0 & -\frac{p_x}{p^2} \\ 0 & 0 \end{array}), \quad
B_1 = (\begin{array}{cc}
 \frac{(f p)_x}{p} & \frac{(f p)_{xx}}{p^2} + 2 f q \\
 -2 f p & -\frac{(f p)_x}{p} 
\end{array})
$$
(and its constant multiples) \CHANGE{of} system~\eqref{ds}, valid without any constraint on $f(t,x)$. We also get a solution of the form shown in Example~\ref{false1}, which corresponds to a false parameter. 

This implies that $H^1$ is always nonzero. This means that all inhomogeneous nonlinear Schr\"o\-dinger equations admit a nonremovable parameter up to the first order.

To see whether the power expansion can be prolonged further, as a next step we consider the $6 \times 6$ matrices $A^{[2]},B^{[2]},C_{(1)}^{[2]},C_{(2)}^{[2]},C_{(3)}^{[2]}$ from the already known blocks
$A_i,B_i,\ C_{(1)i},\ C_{(2)i},\ C_{(3)i}$, $i \le 1$ and the yet unknown blocks $A_2,B_2,\ C_{(1)2},\ C_{(2)2},\ C_{(3)2}$.
Normalizing $A_2,C_{(3)2}$ in the same way as $A_1,C_{(3)1}$ above, we consider the corresponding system~\eqref{ds}. However, this new system turns out to be incompatible unless 
$$
\numbered\label{fxx}
\frac{d^2 f}{dx^2} = 0.
$$
By Proposition~\ref{prop:ext}, condition~\eqref{fxx} is necessary for the initial zero curvature representation~\eqref{nhnls AB} to admit a nonremovable parameter, assuming the jet order at most two.   

{\it Choice 2. }
According to the first row of Table~\ref{tabsl2}, $C_{(3)0}$ is in a normal form, whilst $A_0$ is not. Normalizing $A_0$ increases the jet order of $A_0,B_0$ to two.
This leads us immediately to computations on the second order jet level, while avoiding the ``false'' parameter solution.

\section{Extension procedures. Recursion operators}

The procedure allows us to compute every term of the Taylor expansion of $A(\lambda), B(\lambda)$.
In the case of polynomial dependence on the spectral parameter we get a closed-form result after a finite number of steps. Otherwise we are left with a truncated Taylor expansion, which is, however, not entirely useless. In principle, one can apply the characteristic element method to compute the full zero curvature representation $A(\lambda), B(\lambda)$. Solving the nonlinear system~\eqref{ds} by triangularization~\cite{Hu} usually involves heavy branching, but knowing the truncated Taylor expansion~\eqref{TaylorAB} permits cutting off many branches immediately. 

The symmetry method~\cite{C-ls} and its extension~\cite{C-nls,C-nhnls,C-G-S} by
Cie\'sli\'nski take advantage of linearity of the problem. Remarkably enough, certain aspects of this method admit a cohomological interpretation, which will be discussed elsewhere.

Apart from zero curvature representations, two-dimensional integrable systems usually possess infinite hierarchies of symmetries, generated by recursion operators. 
Given a zero curvature representation or a Lax pair, the recursion operator $\mathcal R$ can be derived by various methods, see~\cite{G-K-S,Sak} and references therein. 
Another relation was observed to exist between the zero curvature representation and the inverse $(\mathcal R + \lambda \Id)^{-1}$~\cite{B-M,M-S,rovee}, which even extends to certain multidimensional systems~\cite{roigsge}. 
\CHANGE{This, however, requires interpretation of the recursion operator as a B\"acklund autotransformation of the linearized equation (Papachristou~\cite{Papa}, Guthrie~\cite{Gut}).}

\CHANGE{Although this interpretation first appeared in multidimensional integrability theory \cite{Papa}, it turned up to be extremely fruitful in two-dimensional case as well.} 
Guthrie's motivating question was how to apply a recursion operator to a symmetry in a safe way. The dangerous point is that the common pseudodifferential form of a recursion operator~\cite{Olv} involves a purely formal inverse 
$D^{-1} = D_x^{-1}$. 
The usual answer is to interpret $H = D_x^{-1}F$ as $F = D_x H$, but this can easily lead to errors (`bogus' symmetries~\cite{Gut}). Guthrie's idea was to provide also the value $G = D_t H$, where $t$ is the other coordinate. This suffices to determine the pseudopotentials up to an integration constant. The integration constant then can be safely omitted since it only adds $\R(0)$ to the result.
Sergyeyev~\cite{Serg} discussed another interpretation problem absent under Guthrie's approach. 

As is well known, infinitesimal symmetries of the system~\eqref{sys} can be put in the vertical form
$$
\numbered\label{ed}
U = \sum_{i,I} D_I U^i \frac{\partial}{\partial u^i_I}, 
$$
where $I$ runs over all multiindices over $t,x$, coefficients $U^i$ being functions defined on $\mathcal E$. The vector field \eqref{ed} is a symmetry if and only if the functions $U^l$ satisfy the system
$$
\numbered\label{ell}
0 = 
\left.
 \sum_{i,I} \frac{\partial E^l}{\partial u^i_I} D_I U^i
\right|_{\mathcal E}
\left.
 =: \ell_{E^l}(U)
\right|_{\mathcal E}.
$$
Formally, \eqref{sys} and \eqref{ell} can be written together as a system of partial differential equations
$$
\numbered\label{linsys}
E^l = 0, \\
\sum_{i,I} \frac{\partial E^l}{\partial u^i_I} U^i_I = 0,
$$
where the components $U^i$ of a symmetry are additional unknowns.
System~\eqref{linsys} will be called the {\it linearized system}.
This allows us to interpret recursion operators as B\"acklund autotransformations for solutions $U^i$ of the linearized system~\eqref{linsys} (Papachristou~\cite{Papa}).

\begin{example} \rm
\label{ex:SW}
Considering the linearized Burgers equation 
$$
u_t = u_{xx} + u u_x, \\
U_t = U_{xx} + u U_x + u_x U,
$$ 
it can be easily checked that the well-known $t$-dependent recursion operator
$$
U' = t U_x + \frac12 (t u + x) U + \frac12 (1 + t u_x) W, \\
W_x = U, \\ 
W_t = U_x + u U,
$$ 
is actually a B\"acklund autotransformation. 
\end{example}

General recursion operators of system~\eqref{sys} are to be sought among B\"ack\-lund autotransformations of the linearized system~\eqref{linsys}, with pseudopotentials subject to first-order linear systems of the form~\eqref{lin problem}. The coefficient matrices $A,B$ are supposed to depend on $u^i_I$ and $U^i_I$ now. Hence the pair $A,B$ constitutes a zero curvature representation over the linearized system~\eqref{linsys}. 

Numerous examples~\cite{Bar,B-M,M-S,rovee,roigsge} (as well as unpublished ones) support the following conjecture:

\begin{conjecture}
\label{conjecture1}
For every recursion operator $\R$ of a system~\eqref{sys} there exists a finite-dimensional Lie algebra $\mathfrak g$ and a zero curvature representation $A,B$ of the system~\eqref{sys} such that the pseudpotentials of $\R$ can be assembled in a single $\mathfrak g$-valued nonlocal variable $\Phi$ subject to conditions
$$
\numbered\label{eq:adcovering}
\Phi_x = [A,\Phi] + \ell_A (U), \qquad
\Phi_t = [B,\Phi] + \ell_B (U).
$$
\end{conjecture}

Here $U$ denotes a seed symmetry, $[\hbox{--},\hbox{--}]$ the commutator in $\mathfrak g$, and $\ell$ the componentwise Fr\'echet derivative 
$$
\ell_A (U) = \sum_{k,I} \frac{\partial A}{\partial u^k_I} U^k_I.
$$

For `conventional' recursion operators the algebra $\mathfrak g$ is typically solvable and mostly abelian, in which case the zero curvature representation $A,B$ reduces to a collection of conservation laws (possibly nonlocal)~\cite{rzcr}.
E.g., in Example~\ref{ex:SW} 
the algebra is 1-dimensional, $u_t = (u_x + \frac12 u^2)_x$ 
being the corresponding conservation law.

On the other hand, the inverse $(\R + \lambda\Id)^{-1}$ of a conventional recursion operator $\R + \lambda\Id$ is usually associated with the $\lambda$-dependent zero curvature representation $A,B$ of system~\eqref{sys} (however, it may happen that the algebra $\frak g$ is solvable).
This is another expression of the conventional wisdom that a recursion operator $\mathcal R$ yields a zero curvature representation through the eigenvalue problem $\R U = \lambda U$.

What we have obtained is a way from a $\lambda$-independent zero curvature representation to a recursion operator $\mathcal R$, the inverse $(\mathcal R + \lambda`Id)^{-1}$, and its associated $\lambda$-dependent zero curvature representation. To make it into a working algorithm we must further restrict the form of the recursion operator.

\begin{conjecture}
For every integrable system~\eqref{sys} there exists a zero order recursion operator with diagonal matrix, i.e., of the form 
$$
\numbered\label{0RO}
\R(U) = (U^{i\prime}) = (c^i_i U^i + a^i_j \Psi^j)
$$
with $\Psi$ as in \eqref{eq:adcovering}, where $c^i_i,a^i_j$ are functions on~$\mathcal E$.
\end{conjecture}

The nonlocalities $\Psi^j$ can take values in an extension of the algebra $\frak g$ of the initial zero-curvature representation. Leaving this aspect aside, we propose the following procedure.

\begin{procedure} \rm
Given an initial zero curvature representation $A = A_0$, $B = B_0$ of~\eqref{sys},
\paritem{1.} write down the linearized system \eqref{linsys} and the associated covering \eqref{eq:adcovering};
\paritem{2.} find all operators \eqref{0RO} such that $U' = \R(U)$ satisfies the linearized system \eqref{linsys}; 
\paritem{3.} solve equation $U' = \R(U) + \lambda U$ for $U$ in terms of $U'$ to obtain the inverse recursion operator $(\R + \lambda`Id)^{-1}$;
\paritem{4.} transform $(\R + \lambda`Id)^{-1}$ to its Guthrie form and identify the associated zero curvature representation (which occurs in adjoint representation here).
\end{procedure}

This procedure was powerful enough to lead to all recursion operators published in the works~\cite{Bar,B-M,M-S,rovee,roigsge}. However, not always a recursion operator could be associated with every value of the spectral parameter (for an instance see~\cite{M-S}). Consequently, for certain $\lambda$-independent zero curvature representations the above procedure fails even though the corresponding $\lambda$-dependent family exists. This phenomenon must be further investigated.

Nevertheless, in the following section we show that our procedure copes well with the zero curvature representation of the inhomogeneous nonlinear Schr\"odinger equation.

\section{The inhomogeneous nonlinear Schr\"o\-dinger equation continued}
\label{sect:nhnls2}

Continuing the example of the inhomogeneous nonlinear Schr\"odinger equation, we consider the nonparametric $\mathfrak{sl}(2)$-valued zero curvature representation \eqref{nhnls AB}  
valid for all functions $f(t,x)$.

{\it Step 1. } The linearized system~\eqref{linsys} is
\begin{eqnarray*}
q_t &=&  (f q)_{xx} + 2 q r, \nonumber
\\
p_t &=&  -(f p)_{xx} - 2 p r, \nonumber
\\
r_x &=&  f (p q)_x + 2 f_x p q, \nonumber
\\
Q_t &=&  (f Q)_{xx} + 2 (r Q + q R), \nonumber
\\
P_t &=&  -(f P)_{xx} - 2 (r P + p R), \nonumber
\\
R_x &=&  f (q P + p Q)_x + 2 f_x (q P + p Q). \nonumber
\end{eqnarray*}
The covering~\eqref{eq:adcovering} is determined by equations 
$$
\numbered\label{nhnls psi}
(\begin{array}{c} 
  \psi_{11} \\ \psi_{12} \\ \psi_{21} 
\end{array})_x
 = (\begin{array}{ccc} 
  \phantom{-}0 & p & q \\ -2q & 0 & 0 \\ -2p & 0 & 0
\end{array})
(\begin{array}{c} 
  \psi_{11} \\ \psi_{12} \\ \psi_{21} 
\end{array}) 
 + (\begin{array}{c} 
  \phantom{-}0 \\ \phantom{-}Q \\ -P 
\end{array}),
\\
(\begin{array}{c} 
  \psi_{11} \\ \psi_{12} \\ \psi_{21} 
\end{array})_t
 = (\begin{array}{ccc} 
  \phantom{-}0 & \phantom{-}(f p)_x & -(f q)_x \\
 \phantom{-}(f q)_x & -2 r & 0 \\ -(f p)_x & \phantom{-}0 & \phantom{-}2 r
\end{array})
(\begin{array}{c} 
  \psi_{11} \\ \psi_{12} \\ \psi_{21} 
\end{array}) 
 + (\begin{array}{c} 
  0 \\ (f Q)_x \\ (f P)_x 
\end{array}),
$$
where $\psi^{ij}$ are components of an $\mathfrak{sl}(2)$-matrix 
$$
\Psi = (\begin{array}{cr} 
  \psi_{11} & \psi_{12} \\ \psi_{21} & -\psi_{11} 
\end{array}).
$$

{\it Step 2. } The equations to be solved are
$$
\numbered\label{eq:ROnhnls}
D_t Q'
 - f D_{xx} Q'
 - 2 f_x D_x Q'
 - (f_{xx} + 2 r) Q'
 - 2 q R' = 0, 
\\
D_t P'
 + f D_{xx} P'
 + 2 f_x D_x P'
 + (f_{xx} + 2 r) P'
 + 2 p R' = 0, 
\\
D_x R'
 - f q D_x P'
 - (f q_x + 2 f_x q) P'
 - f p D_x Q'
 - (f p_x + 2 f_x p) Q' = 0
$$
where, according to~\eqref{0RO}, we assume
$$
P' = c_P P + a_P^{11} \psi_{11} + a_P^{12} \psi_{12} + a_P^{21} \psi_{21}, \\
Q' = c_Q Q + a_Q^{11} \psi_{11} + a_Q^{12} \psi_{12} + a_Q^{21} \psi_{21}, \\
R' = c_R R + a_R^{11} \psi_{11} + a_R^{12} \psi_{12} + a_R^{21} \psi_{21}. \\
$$
Functions $c_P,c_Q,c_R$, $a_P^{ij},a_Q^{ij},a_R^{ij}$ are the unknowns and can depend on $t,x,\ p,q,r$ and their derivatives $p_x,q_x,r_t$, $p_{xx},q_{xx},r_{tt},\dots$. 
System~\eqref{eq:ROnhnls} can be solved routinely. As a result we obtain that $\R(U)$ is a constant multiple of $U$ unless $f_{xx} = 0$ again. On the other hand, if $f = f_1(t) x + f_0(t)$, then we obtain a nontrivial solution
$$
\numbered\label{nhnls RO}
P' = g_1 P + \psi_{21}, \\
Q' = g_1 Q + \psi_{12}, \\
R' = g_1 R + f_1 \psi_{11} + f p \psi_{12} + f q \psi_{21}, 
$$
where
$$
\numbered\label{g1}
g_1 = \int f_1(\tau)\,d\tau. 
$$

The recursion operator given by formulas~\eqref{nhnls RO} and~\eqref{nhnls psi} can be applied to an arbitrary seed symmetry. However, generated symmetries will be nonlocal as a rule, which is usually the case with inverse recursion operators. 

{\it Step 3. } To invert this operator, we reinterpret formulas~\eqref{nhnls RO} and~\eqref{nhnls psi} so that $P',Q',R'$ are given and $P,Q,R$ are to be found. 
From equation~\eqref{nhnls RO} we get
$$
\numbered\label{nhnls iRO}
P = \frac{P' - \psi_{21}}{g_1}, \
Q = \frac{Q' - \psi_{12}}{g_1} , \
R = \frac{R' - (f_1 \psi_{11} + f p \psi_{12} + f q \psi_{21})}{g_1}, 
$$
while $\psi_{ij}$ will satisfy a system of the same form~\eqref{eq:adcovering}. 
The term $\lambda`Id$ is absorbed in the integration constant of~\eqref{g1}. 
Herefrom we can reconstruct the corresponding zero curvature representation as
$$
\numbered\label{nhnls lambda}
A(\lambda) = (\begin{array}{cc} 
  -\frac1{2 g_1} & q \\ -p & \frac1{2 g_1} \end{array}), \quad
B(\lambda) = (\begin{array}{cc} 
  r + \frac{f}{2 g_1^2} & (f q)_x - \frac {f q}{g_1} \\ 
  (f p)_x + \frac {f p}{g_1} & -r - \frac{f}{2 g_1^2} \end{array}),
$$
where $g_1$ is the integral~\eqref{g1} and the integration constant for $g_1$ serves as the spectral parameter $\lambda$. This result is gauge equivalent to that obtained by Cie\'sli\'nski~\cite{C-nls,C-nhnls}.

That the parameter is nonremovable can be seen from the expansion of $A$ in powers of $\lambda$. Indeed, $A_1\,dx + B_1\,dy$ is then gauge equivalent to the previously computed generator~\eqref{nhnls H^1} of $H^1$.  
In particular, the upper and lower bounds established in Sections~\ref{sect:nhnls1} and~\ref{sect:nhnls2} coincide, meaning that the answer is complete
(within the Lie group $\mathfrak{sl}(2)$ and the second jet order of the zero curvature representation).

Remarkably enough, the inverse of the recursion operator~\eqref{nhnls iRO} still fails to produce a local hierarchy.

\section*{Acknowledgements}

The author owes very much to I.S Krasil'shchik (most of the inspiration I got from his work~\cite{Kra,K-K}) and J. Cie\'sli\'nski.
The support from M\v{S}MT project MSM 4781305904 is gratefully acknowledged.

\end{document}